\documentclass[12pt]{iopart}
\input{amssym}
\usepackage{epsfig}
\global\arraycolsep=2pt

\newcommand{\be}{\begin{equation}}

\newcommand{\ee}{\end{equation}}
\newcommand{\bea}{\begin{eqnarray}}
\newcommand{\eea}{\end{eqnarray}}
\newcommand{\ben}{\begin{equation*}}
\newcommand{\een}{\end{equation*}}
\newcommand{\bean}{\begin{eqnarray*}}
\newcommand{\eean}{\end{eqnarray*}}

\begin{document}
\title{Decoherence in a Two Slit Diffraction Experiment with Massive Particles
}
\author{Paula I. Villar and Fernando C. Lombardo}
\address{Departamento de F\'\i sica {\it Juan Jos\'e
Giambiagi}, Facultad de Ciencias Exactas y Naturales, UBA; Ciudad
Universitaria, Pabell\' on I, 1428 Buenos Aires, Argentina}

\ead{paula@df.uba.ar}

\begin{abstract}
Matter-wave interferometry has been largely studied in the last few
years. Usually, the main problem in the analysis of the
diffraction experiments is to establish the causes for the loss of
coherence observed in the interference pattern. In this work, we
use different type of environmental couplings to model a two slit
diffraction experiment with massive particles. For each model, we
study the effects of decoherence on the interference pattern and
define a visibility function that measures the loss of contrast of
the interference fringes on a distant screen. Finally, we apply our 
results  to the experimental reported data on massive particles 
$C_{70}$.

\end{abstract}


\newcommand{\beq}{\begin{equation}}
\newcommand{\eeq}{\end{equation}}
\newcommand{\dalam}{\nabla^2-\partial_t^2}
\newcommand{\mbf}{\mathbf}
\newcommand{\itm}{\mathit}
\newcommand{\beqa}{\begin{eqnarray}}
\newcommand{\eeqa}{\end{eqnarray}}

\vspace{0.5cm}
 The wave-particle duality of material objects is a hallmark of
quantum mechanics. Up to now, the wave nature of particles has
been demonstrated for electrons, neutrons, atoms, and coherent
atomic ensembles. Yet more important, many theoretical studies
have been done around the mesoscopic systems
\cite{Facchi:2005,Brezger:2002}. Mesoscopic objects are neither
microscopic nor macroscopic. They are generally systems that can
be described by a wavefunction, yet they are made up of a
significant number of elementary constituents, such as atoms.
Well-known examples these days are fullerene molecules $C_{60}$
and $C_{70}$, which are expected to behave like classical
particles. Nonetheless, the quantum interference of these
molecules has been observed \cite{Zeilinger:2003}. In particular,
the advances with large molecules have stimulated the question
what determines the limits to observe quantum delocalization with
massive objects. Thus, there is a need to theoretically quantify
the effect of decoherence (or dephasing) on the observed
interference pattern in a double-slit experiment. It is quite
intuitive that the resulting pattern shall be an interplay between
the strength of the coupling to the environment, the slit
separation and the distance the particle travels from the slit to
the screen.

We shall study a two slit diffraction experiment with particles of
mass M diffracted by a grating (in the x direction) and then
detected on a distant screen (located a distance L in the y
direction). Note that coherence in the x direction is required in
order to observe an interference pattern on the screen, whereas the
dynamics in the y direction can be that of a free non-interacting
particle since it just serves to transport the particle from the
grating to the screen. Initially, we may reasonably assume that the
action of the grating is to prepare a superposition of two Gaussian
wavepackets (which best describe a massive particle), centered at
each location of the respective slits and factorized as
\cite{Venugopalan:2006,Viale:2003} $
\Psi(\vec{x},0)= (\phi_1(x,0) + \phi_2(x,0))  \chi(y,0) \otimes
\zeta(\vec{X},0),$
where $|\phi_1|^2$ and $|\phi_2|^2$ correspond to the probability amplitudes
for the particle to pass through slit $1$ and slit $2$ (in the
x-axis), respectively, while $\chi(y,t)$ represents the Gaussian
wave function in the y direction (where no superposition is needed)
and $\zeta(\vec{X},t)$ describes the state of the environment
coupled to the subsystem. Note that we are assuming translational
invariance in the z-axis \cite{Viale:2003}.

We shall consider the environment as a set of non-interacting
harmonic oscillators and the dynamics of the particles modeled by a
quantum brownian motion (QBM) \cite{Hu:1992}.  This behavior can be
imputed to the passing of the particles through the slits producing
vibrations or any other kind of interactions with the walls of the
grating able to corrupt the interference pattern. Moreover, also the
finite size of the grating and the differences in the slit apertures
can attenuate the visibility of the interference fringes, especially
in the case of complex molecules such us $C_{60}$ and $C_{70}$
\cite{Horn:2004}. In order to study the interference pattern
registered on the screen at a later time $t_L$, we need to obtain
the evolution in time of the reduced density matrix
$\rho_r(x,x',t)$, which is given by the following master equation
\beqa \frac{\partial \rho_r(x,x',t)}{\partial t}& =& \frac{i
\hbar}{2 M} \bigg( \frac{\partial^2 \rho_r}{\partial x^2}-
\frac{\partial^2 \rho_r} {\partial x'^2}\bigg) - \frac{{\cal
D}(t)}{4 \hbar^2} (x-x')^2 \rho_r \nonumber \\
&-& \gamma(t) (x-x') \bigg(\frac{\partial \rho_r}{\partial x}-
\frac{\partial \rho_r}{\partial x'} \bigg) + 2 f(t) (x-x')
\bigg(\frac{\partial \rho_r} {\partial x} + \frac{\partial
\rho_r}{\partial x'}\bigg), \label{master} \eeqa where  $\gamma(t)$
is the dissipative coefficient (proportional to the square of the
coupling constant to the environment), ${\cal D}(t)$ the diffusive
coefficient and $f(t)$ the coefficient responsible for the anomalous
diffusion. Eq.(\ref{master}) has been obtained by assuming the
environment to be in equilibrium, at a temperature T.  In the case
that the system is coupled to an ohmic environment in the high
temperature limit ($k_B T >> \hbar \omega$), these coefficients are
constant $\gamma(t)=\gamma_0$, ${\cal D}(t) = 2 M \gamma_0 k_B T$
and $f(t) \approx 1/k_B T$ in units of $\hbar=1$
 \cite{Hu:1992}. It is important to stress
that the equation of movement for the  generally used scattering
model $ i \frac{\partial \rho_r}{\partial t} = [H,\rho_r] - i
\Lambda [x,[x,\rho_r]]$, where the effect of the environment is
included in the collision term $\Lambda$, can be obtained from
Eq.(\ref{master}) in the high temperature limit (neglecting
dissipation) for the markovian case if written in the Lindblad
form. However, this master equation refers to a more general
movement that can be used for all temperatures, even to study
 the dynamics of the
test particle  at zero temperature. Yet more interesting, this
formulation includes the phenomenological scattering model and verifies the
fluctuation-dissipation theorem for a general system in thermal
equilibrium.

The interference pattern corresponds to the probability distribution
of the time evolved wave function: \beqa
P(\vec{x},t)&=&|\Psi(\vec{x},t)|^2=\bigg( \phi_1(x,t)^* \phi_1(x,t)+
\phi_2(x,t)^* \phi_2(x,t) \nonumber \\
&+&\phi_2(x,t)^* \phi_1(x,t) 
+\phi_1(x,t)^* \phi_2(x,t) \bigg) \otimes  \chi(y,t)^* \chi(y', t),
\nonumber  \eeqa which is proportional to the
diagonal part of the density matrix defined as
$\rho(\vec{x},\vec{x}',t)= |\Psi(\vec{x},t)\rangle \langle
\Psi(\vec{x}',t)|$. In the case we are studying herein, i.e an open
quantum system, the interference pattern at a given time $t$ on the
screen is given by: \[ P(\vec{x},t)=
\bigg(|\phi_1(x,t)|^2+|\phi_2(x,t)|^2 
+ 2 {\bf \Gamma}(t){\rm Re} (\phi_1^*(x,t)
\phi_2(x,t)) \bigg) |\chi(y,t)|^2,  \label{gamma} \] where  
the terms in brackets correspont to $\rho_r(x,x,t)$  and ${\bf
\Gamma}(t)$ encodes the information about the statistical nature of
noise since it is obtained after tracing out the degrees of freedom
of the environment. ${\Gamma}(t)$ is an exponential 
decaying term which
suppresses the interference terms in a decoherence time scale $t_D$.
In the case of the QBM, $\Gamma(t)=e^{- \Delta x^2 
\int_0^t {\cal D}(s) {\rm ds}}$ (with $\Delta
x^2=(x-x')^2$)
 represents the
noise induced environmental effect on the system. \begin{figure}
\centerline{\psfig{figure=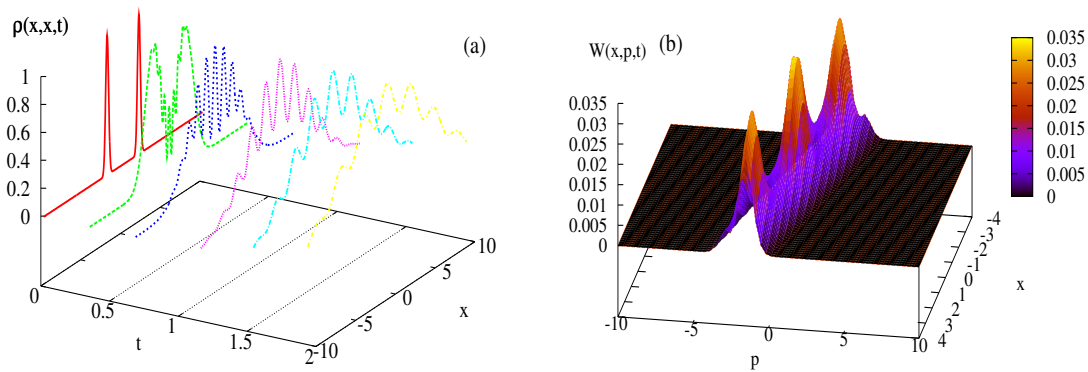,height=6cm,width=15cm,angle=0}}
\caption{(a) Evolution in time for the interference pattern registered
on the screen when the system is coupled to an environment.
Parameters used: $L_0=2$, $\sigma_{x0}=0.5$, $M=1$,
$\gamma_0=0.001$ and $k_BT=300$. (b) For the same values the wigner
function at $t=2~{\rm s}$ shows that the interferences have
disappeared since the wigner is positive.} \label{fig1}
\end{figure}
In this framework, we numerically solved the master equation
Eq.(\ref{master}) for the reduced density matrix $\rho_r(x,x',t)$ of
two well localized Gaussian wave packets initially given by \beqa
\Psi(\vec{x},0)&=& N \bigg( \exp(\frac{(x-L_0)^2}{4 \sigma_{x0}^2})
+ \exp(\frac{(x+L_0)^2}{4 \sigma_{x0}^2}) \bigg) \otimes
\exp(-\frac{y^2} {4 \sigma_{y0}^2}-i k_y y), \nonumber \label{2gaus}
\eeqa where $2L_0$ is the initial separation of the center of the wave packets,
$\sigma_{x0}^2$  and
 $\sigma_{y0}^2$ are the initial
width of the packet in the x and y-axis, respectively, and $k_y$ the
initial moment of the particle in the y direction. It is important
to note that $L_0$, $\sigma_{x0}$, $\sigma_{y0}$ and $k_y$ are all
free parameters that have to be tuned with the experimental data. In
addition, we assume that $\Delta p_y << p_y$, so the moment
component is sharply defined and the wave packet has a
characteristic wavelength $\lambda_{dB}$ associated $\lambda_{dB}
\sim \hbar/p_y << \Delta y$ \cite{Sanz, tumulka}.
 Once the reduced density matrix is
known for all time $t$, we know the dynamics of these two packets
and can study the effects of decoherence on the interference pattern
registered on the distant screen. In Fig.\ref{fig1}(a) we present
the time evolution  for the interference pattern registered on the
screen. We can there observe the two wave packets initially
separated a distance $2L_0$ which start to spread as time increases.
In a short timescale, interferences start to develop, however we can
clearly observe that the minimum fringes of the interference pattern
are not exactly zero. This means that there is a loss of contrast
between theis case (an open system) with respect to the interference
pattern of an unitary evolution. This means that decoherence is
already playing a crucial role in determining the wave behavior of
the massive particle. In Fig.\ref{fig1}(b) we  present the wigner
function for the corresponding reduced density matrix at a fixed time
 $t=2~{\rm s}$. We can see that it is a completely positive function, which
assures us that decoherence has been effective in this short
timescale. We shall estimate the decoherence time $t_D$, i.e. the timescale for
which the interferences are mostly destroyed (up to a $70~\%$)
 as $\Gamma_{{\cal
D}}(t_D)=\exp(-{\cal D}\Delta x^2 t) \sim 1/e$.  It is easily
deduced that $t_D \approx 1/({\cal D} \Delta x^2)$. A
quantity of particular importance in matter wave
 interferometry is
the fringe visibility $\nu(t)$. In order to study the
 role of decoherence on the
 fringe visibility, we define it as
$ \nu(t) \sim \frac{|\rho_{\rm int}(x,x,t)|}{\rho_{11}(x,x,t)
 + \rho_{22}(x,x,t)} = \frac{\Gamma(t)}{\rho_{11}(x,x,t)
 + \rho_{22}(x,x,t)} $ where
$\rho_{ii}=|\phi_i(x,t)|^2$, with $i=1,2$ and $\rho_{\rm{int}}$ the
interference terms.
\begin{figure}
\centerline{\psfig{figure=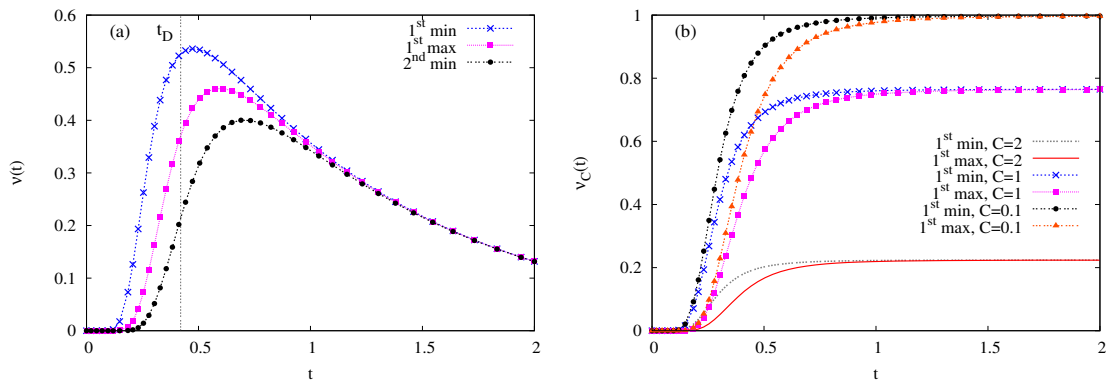,height=6cm,width=15cm,angle=0}}
\caption{(a) Time evolution for the  visibility fringe
$\nu(t)$ for $k_BT=300$, $\gamma_0=0.001$, $L_0=2$, $M=1$ and
$\sigma_{x0}=0.5$.  The estimation of the decoherence time $t_D \sim
1/(2 M \gamma_0 k_BT L_0^2)=0.41~s$ coincides with the timescale at
which the visibility starts to decrease towards a null value. (b)
 Time evolution for the visibility function $\nu_C(t)$ for
neutrons ($C_{\rm N}=0.1$) and fullerenes ($C_{\rm
F}=1$ and $C_{\rm F}=2$ ) in the presence of an external
 classical  time dependent
electromagnetic field (incoherence effects). 
The curves  are for the first minimum and
maximum of the interference pattern and reach a different asymptotic
value than in the case of $\nu(t)$. } \label{fig2}
\end{figure}

Clearly, the visibility fringe goes down as $t_L$, i.e. the
observation time, is larger than the decoherence time $t_D$.
However, if we succeed in performing our two slit experiment in a
time $t_L < t_D$ at a fixed room temperature $k_BT$, we can see that
the visibility fringe decreases as the coupling to the environment
 ($\gamma_0$) increases. This is so, because
the decoherence time depends inversely on the coupling constant:
the bigger $\gamma_0$, the shorter the decoherence time. Not only
can we check the dependence upon the coupling constant but also its
time evolution. This behavior is shown in Fig.\ref{fig2}(a). For
short times, the visibility increases from zero to a maximum value
because the interferences start to develop at that short timescale
but are not present at $t=0$ (since the wave packets are initially
separated and have to spread so as to generate the interferences see
Fig.\ref{fig1}(a)). This maximum value coincides with the estimated
decoherence time $t_D$. Then, the visibility starts to decrease,
since the destruction of the interferences is taking place. Note
that the visibility is a quantity that measures the loss of contrast
of the interference fringes. Then, it is expected that those with
the bigger contrast suffer from this attenuation the more, as seen
in Fig.\ref{fig2}(a). Clearly, the observation time $t_L$ must be
shorter than the decoherence time in order to observe the
interference pattern.

It shall also be interesting to study the visibility function 
for a model of incoherence such as the one presented by us 
in \cite{Nos}. In particular,
$\Gamma_C=J_0(|C|)$ is constant in time
for the experimental data of neutrons and fullerenes \cite{Nos}. Therein, we
previously estimated the quantity $C$ for these massive particles
and observed that, contrary to might be naively expected, in thought
and real experiments, $C_{\rm F} \sim {\cal O}(1)$.  However, 
for neutral particles with
permanent dipole moment this value is much lower $C_{\rm N}
\sim {\cal O}(0.01)-{\cal O}(0.1)$. Therefore, we define the visibility
 function $\nu_C(t)$
as $ \nu_C(t)=\frac{J_0(|C|)}{\rho_{11}(x,x,t)
 + \rho_{22}(x,x,t)}$ and show its time evolution for neutrons 
 and fullerenes in Fig.\ref{fig2}(b). We recall that
this interaction is always present in charged and neutral particles
with permanent dipole moment and can never be turned off. However,
in the case of neutrons it has been shown in \cite{Nos} that it can
be neglected, whereas it is unexpectedly important for massive
particles such as fullerenes.
 \begin{figure}
\centerline{\psfig{figure=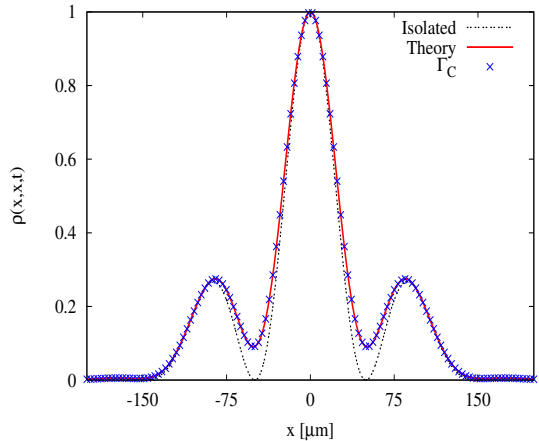,height=6cm,width=7.5cm,angle=0}}
\caption{The interference pattern ($\nu \sim 0.62$) registered on
the screen considering the incoherence (or dephasing) effects to model  a two slit
diffraction experiment with massive particles $C_{70}$. The curves
are for the isolated case, the theoretical prediction using the experimental 
data  reported in \cite{Zeilinger:2003} and
the incoherence model using $\Gamma_C$ with $C=1$.} \label{fig3}
\end{figure}

Finally, we shall apply our models of decoherence and incoherence 
to the experimental
data reported in \cite{Zeilinger:2003} to reproduce the observed
 pattern for fullerenes $C_{70}$. In Fig.\ref{fig3}, the
 interference pattern  is shown. Therein, we
have considered the unitary and nonunitary evolution  of the particles.
 For these massive particles, we can see that the interference pattern 
is attenuated when the system is open. 
What is more important, is that the modeling of the incoherence effects 
through the overlap factor $\Gamma_C$ is  enough to reproduce the
effects of the environment on the interference pattern of
 a real diffraction experiment with massive particles.

All in all, we have presented a fully quantum mechanical treatment
using a microscopic model of environment and also a concrete example
to include the incoherence effects. Therefore, we have studied the
effects of decoherence on the interference pattern of thought
diffraction experiments and presented an analysis of matter wave
interferometry in the presence of a dynamical quantum environment
such as the quantum brownian motion model. We have shown the
interference patterns and visibility function $\nu(t)$ for thought
diffracted free particles  in the high temperature limit
(assumption valid for massive particles diffracted at room
temperature). Yet more important, we have defined the visibility 
function $\nu_C(t)$ for a
model environment previously developed which describes the
incoherence effects originated in the experimental difficulty of
producing the same initial/final state for all particles (i.e the
existence of a random variable such as the particle's emission
time). We showed that it is qualitatively different than the one
commonly found in the literature and very important in the case of
diffraction experiments with massive particles such as fullerenes.
In this case, the incoherence effects are enough to model the
attenuation of the interference pattern observed in the real
experiment, whereas in the case of cold neutrons  the incoherence
effects are not of such importance \cite{Nos2}. Therefore, in the latter case we
must consider the decoherence effects by using the corresponding
formulation (such us QBM). This result might have been expected since the
interaction of more massive particles with the external classical
field is more important than for those with a smaller mass where
other kind of interactions (such as with the walls of the grating or
the air molecules) seem to prevail.

We would like to thank  H. Thomas Elze for the organization 
of  DICE'06. This work was supported by UBA, CONICET, 
Fundaci\'on Antorchas, and
ANPCyT, Argentina.  P.I.V.
gratefully acknowledges financial support of UIPAP.

\end{document}